\documentclass[journal=nalefd,manuscript=article]{achemso}

\usepackage[version=3]{mhchem} 

\title{Spin dependent quantum interference in non-local graphene spin valves}

\author{M. H. D. Guimar\~aes}
\email{m.h.diniz.guimaraes@rug.nl}
\author{P. J. Zomer}
\author{I. J. Vera-Marun}
\author{B. J. van Wees}
\affiliation{Physics of Nanodevices, Zernike Institute for Advanced Materials, University of Groningen, The Netherlands}

\begin{document}

\begin{abstract}
Spin dependent electron transport measurements on graphene are of high importance to explore possible spintronic applications.
Up to date all spin transport experiments on graphene were done in a semi-classical regime, disregarding quantum transport properties such as phase coherence and interference.
Here we show that in a quantum coherent graphene nanostructure the non-local voltage is strongly modulated.
Using non-local measurements, we separate the signal in spin dependent and spin independent contributions.
We show that the spin dependent contribution is about two orders of magnitude larger than the spin independent one, when corrected for the finite polarization of the electrodes.
The non-local spin signal is not only strongly modulated but also changes polarity as a function of the applied gate voltage.
By locally tuning the carrier density in the constriction we show that the constriction plays a major role in this effect and indicates that it can act as a spin filter device.
Our results show the potential of quantum coherent graphene nanostructures for the use in future spintronic devices.

\end{abstract}
\maketitle
\section{Introduction}
Graphene has attracted a lot of attention in the field of spintronics due to theoretical predictions of long spin relaxation length ($\lambda_{s}$) and spin relaxation time \cite{PRL-Brataas}.
Experimentally, although not matching the initial theoretical expectations, graphene has already shown spin information transfer over long distances at room temperature\cite{PRB-Paul} and electrical creation of a large spin imbalance ($\mu_{s}\approx$ 1 meV)\cite{NatPhys-Ivan}.
Despite several works focused on the experimental limits on the spin relaxation in graphene\cite{PRB-Paul,NanoLett-Marcos,PRB-Csaba,PRB-Thomas-Contact,PRB-Thomas-FLG,PRL-Kawakami,JMMM-Kawakami,NanoLett-Kawakami,PRB-Kawakami}, none have shown the effect of quantum transport properties of the charge carriers on the spin dependent transport.
In order to study such effects we have to move away from the semi-classical regime and adopt a quantum mechanical approach.
For this, the device dimensions have to not only be comparable to $\lambda_{s}$ but also to the phase-coherent length ($\lambda_{\phi}$).
Graphene has the advantage that both these characteristic lengths are in the order of a few micrometers \cite{PRB-Falko,JOP-Mason,PRL-Folk-2013,Nature-Niko,PRB-Mihai,PRB-Csaba,PRL-Kawakami}, making the device fabrication for this kind of devices easier than when using regular metals or semiconductors.

The combined effects of confinement, coherence and spin of the charge carriers has led to several ground breaking works on quantum dots \cite{PRL-Folk,Science-Folk,NatPhys-Lindelof,NatPhys-Schonenberger} and other quantum coherent structures such as Fabry-Perot interferometers\cite{NatPhys-Folk,PRB-Mopurgo,PRB-Schonenberger}.
When the electronic transport is studied in a device which dimensions are smaller or comparable to $\lambda_{\phi}$, the conductance shows non-periodic oscillations as a function of the Fermi energy and perpendicular magnetic field.
These oscillations, called universal conductance fluctuations (UCF), are due to quantum interference between the different paths the carriers take when they traverse the device, in a similar way to that of weak-localization (WL) \cite{JOP-Mason,SSP-Beenaker}. 
The interference pattern of the carriers in the device is influenced by the relative phase between different paths.
The relative phase of the carriers depends on their Fermi wave vector and the Aharonov-Bohm flux through the sample.
Particularly for graphene, UCF and WL have provided information on the spin behaviour by using non-magnetic contacts and large in-plane magnetic fields or by studying the temperature dependence of $\lambda_{\phi}$ \cite{PRB-Falko,NatPhys-Folk,PRL-Folk-2013}.

In this letter we demonstrate that a strong spin dependent transmission can arise in a graphene nanodevice when the quantum interference pattern shown as UCF and electrical spin injection and transport are combined.
Using ferromagnetic electrodes we create a spin accumulation that can be quantified by the difference in the chemical potentials for spin up (majority) and spin down (minority): $\mu_{s}=\mu_{\uparrow}-\mu_{\downarrow}$.
When a current is driven in the device and a voltage is measured outside the current path, we can observe oscillations in this non-local voltage as a function of a gate voltage that have charge and spin contributions.
The fact that we use ferromagnetic electrodes allows us to separate the charge and spin contributions showing that the spin dependency in the oscillations is about two orders of magnitude larger than that of the charge when the polarization of the electrodes is taken into account.
We show that the spin signal can be modulated by orders of magnitude and even reverse polarity using only an applied gate voltage to change the Fermi level, indicating that the device can be used as a spin filter without the need of an external magnetic field.

A similar spin filtering effect was already observed in open quantum dots fabricated on a GaAs/AlGaAs 2DEG \cite{Science-Folk}, but relied on a large in-plane magnetic field to create a Zeeman spin splitting, whereas our device operates at zero magnetic field.
It has also been shown that a control of magnitude and reversal of the non-local spin signal in graphene can be obtained using Fabry-Perot cavities \cite{APL-Fuhrer}, but the signals could only be modulated by one order of magnitude or less.
A large oscillating non-local voltage was observed in a series of quantum dots \cite{PRB-Schonenberger}, but the spin dependent nature of the signal could not be measured.
Here we explicitly show that, using a simple device geometry with ferromagnetic contacts, the transmission through the device become strongly spin dependent and can be controlled by a gate voltage.

\section{Methods}
Our samples were obtained by mechanical exfoliation of highly oriented pyrolytic graphite on 500 nm SiO$_{2}$/Si substrates.
Single layer graphene flakes were selected using optical contrast and the flake structure was defined by electron-beam lithography (EBL) followed by reactive ion etching in pure oxygen plasma.
In order to remove polymer remains and keep the graphene-contact interface clean, the samples were heated to 350 $^{o}$C in Ar/H$_{2}$ gas flow for 2 hours.
The electrical contacts to graphene are then made using a second EBL step.
For the contact deposition a 0.4 nm layer of Ti is evaporated in an e-beam evaporator at high-vacuum atmosphere.
In order to fully oxidise the Ti layer and obtain a highly resistive contact interface to avoid the conductivity mismatch problem\cite{PRB-Thomas-Contact}, pure oxygen gas is let in the chamber and the sample is kept in a pressure above 10$^{-1}$ mbar for 15 minutes.
The entire evaporation and oxidation process is repeated once more followed by the evaporation of 35 nm of Co.

\section{Results}
\begin{figure}[h!]
    \centering
        \includegraphics[width=0.5\textwidth]{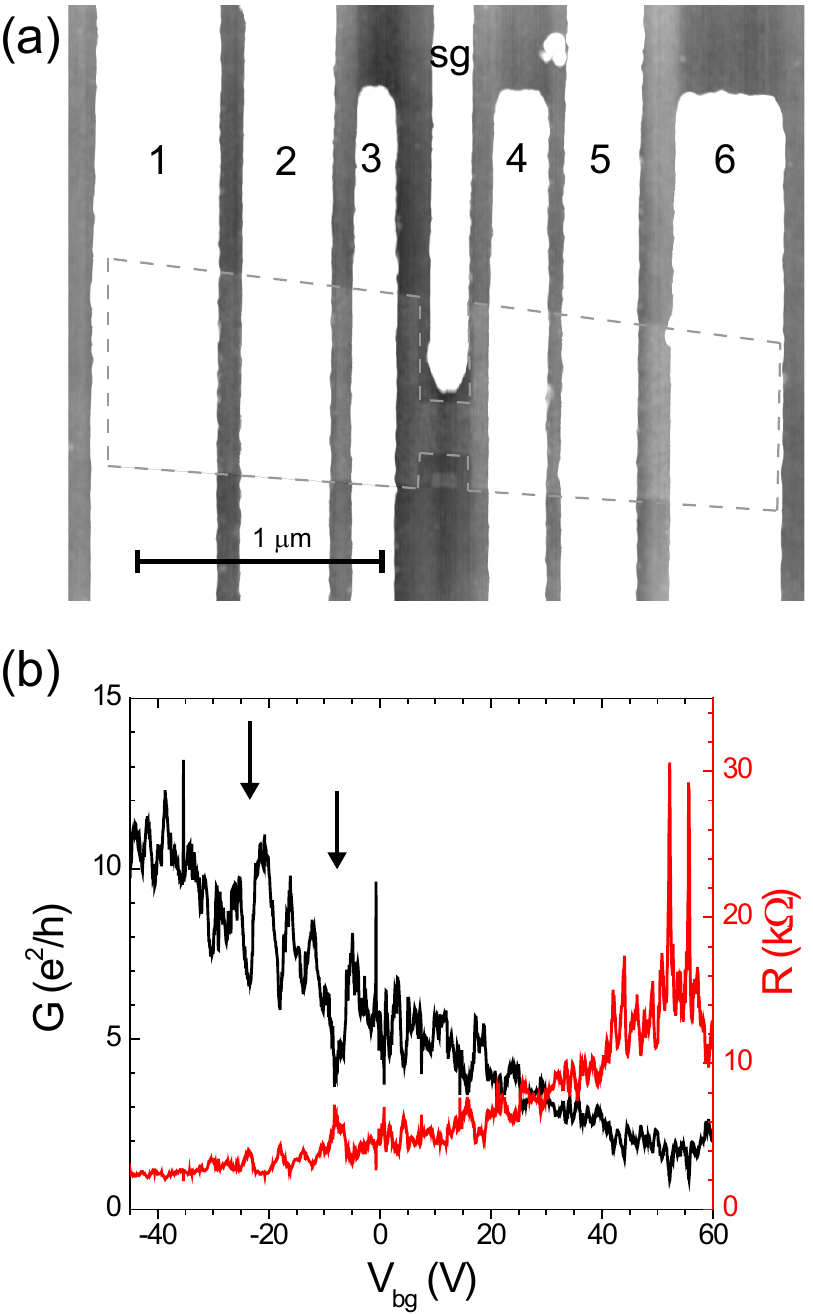}
    \caption{(a) Atomic force micrograph of the device on which the measurements are performed. The graphene structure is outlined by a dashed line for clarity. The contacts numbered from 1 to 6 and the side-gate (sg) are shown in white. (b) The 4-terminal conductance, G (black) and resistance, R (red) as a function of the back-gate ($V_{bg}$) with the side-gate at $V_{sg}$=0 V.}
    \label{fig:fig1}
\end{figure}

Figure \ref{fig:fig1}(a) shows an atomic force micrograph of the sample in which the measurements in this paper were performed.
The device consists of two graphene areas of 1.2 x 0.75 $\mu$m$^{2}$ connected via a 0.2 x 0.25 $\mu$m$^{2}$ constriction.
Three contacts were deposited in each of the wide graphene areas and a side-gate about 100 nm away from the constriction was used to locally control the carrier density.
Similar devices without the constriction in the centre were also measured but did not show the strong modulation on the spin-valve signal as shown here.

All measurements were performed using standard low-frequency lock-in techniques with currents up to 100 nA at a temperature of 4.2 K.
In order to avoid contributions from the contact resistance, all the charge transport measurements were obtained in a local 4-probe configuration.
The spin dependent measurements were performed in a non-local 4-probe geometry \citep{Nature-Niko,PRB-Mihai} to avoid charge transport contributions as described below.
Since our contacts are non-invasive, with contact resistances in the range of 50-200 k$\Omega$, contacts which are not connected do not affect the spin or charge transport measurements.

The local charge transport measurements were performed by applying a current between the outer contacts (1 and 6) and measuring the voltage drop between contacts 2 and 4.
We observe reproducible non-periodic oscillations in conductance (G) as a function of back-gate voltage (V$_{bg}$), Fig. \ref{fig:fig1}(b)).
These oscillations are attributed to UCF. 

In order to separate the spin from the charge contribution to the signal we use a 4-probe non-local technique\cite{Nature-Niko,PRB-Mihai}.
A charge current is driven between contacts 2 and 1 creating a spin accumulation which diffuses away from the point of injection and can then be detected by measuring the voltage difference, V$_{nl}$, between contacts 4 and 6.
Since the charge current path is separated from the voltage detection circuit, most of the charge contribution is in principle excluded.
However, as shown later, due to the quantum coherent nature of the transport a sizeable charge contribution to the non-local signal can be observed.
The non-local voltage can  be normalized as a function of the charge current to obtain a non-local resistance: $R_{nl}=I/V_{nl}$.

\begin{figure}[h]
    \centering
        \includegraphics[width=1.0\textwidth]{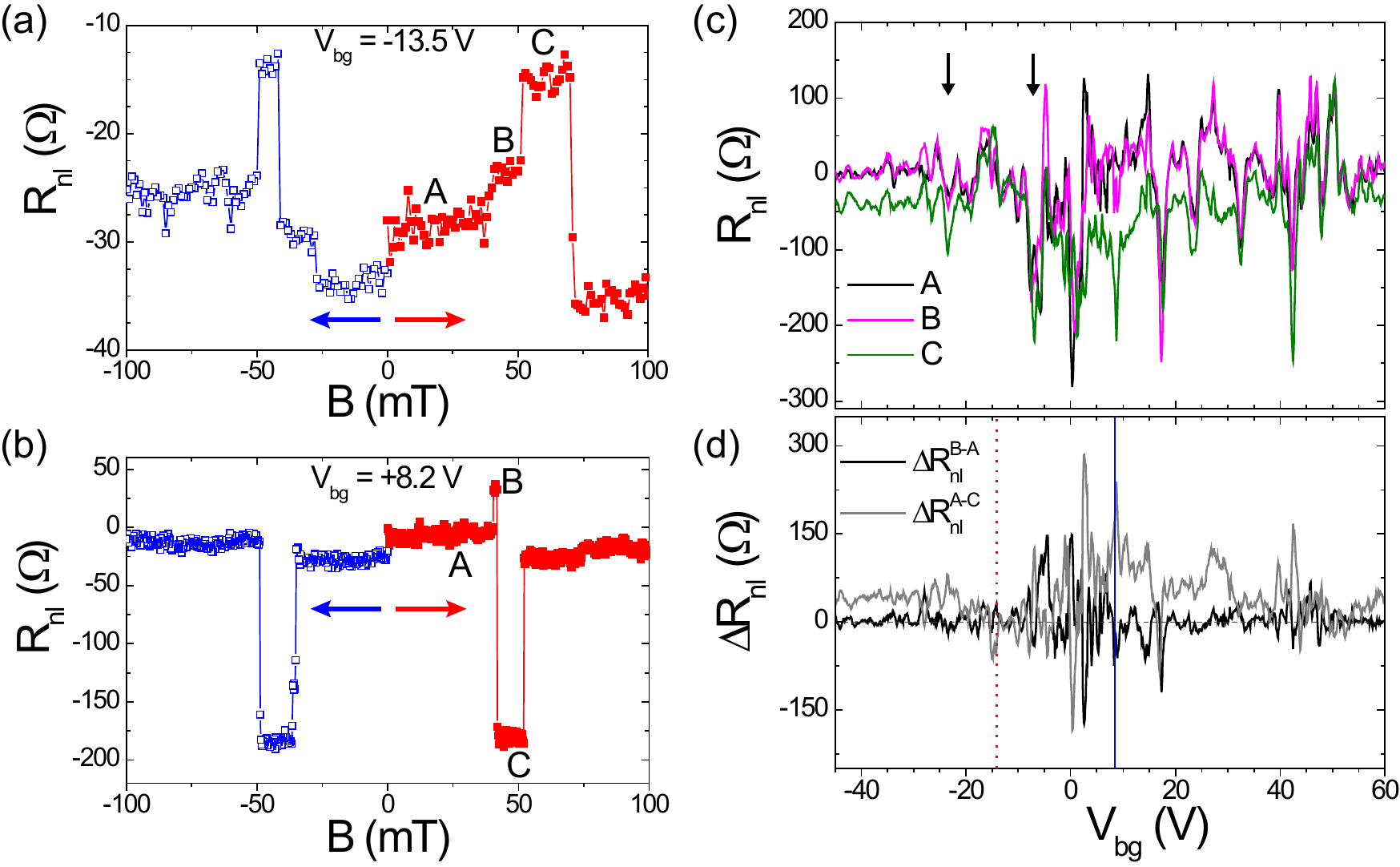}
    \caption{Non-local spin valve measurements at different back-gate voltages: (a) V$_{bg}$=-13.5 V and (b) V$_{bg}$=+8.2 V. The solid red (open blue) points were recorded for a magnetic field sweep from zero to positive (negative) values. The arrows indicate the scan directions. The letters (A, B and C) mark the three levels in the non-local resistance corresponding to different magnetization configuration of the electrodes. (c) The non-local signal as a function of back-gate voltage corresponding to the three magnetization configuration of the contacts: A (black), B (magenta) and C (green). The region in between the arrows should be compared to the region within the arrows in Fig. \ref{fig:fig1}. For these measurements the magnetic field was set to zero after setting the magnetization configuration of the electrodes. (d) Difference between the traces B and A (black) and C and A (grey) showing the spin dependent part of the signal. The red dotted and blue solid lines indicate respectively the values of $V_{bg}$ in which the measurements of (a) and (b) were performed. For all measurements in this figure the side-gate voltage was set to $V_{sg}=0$ V.}
    \label{fig:fig2}
\end{figure}

When a large negative magnetic field (B $\approx$ -1 T) parallel to the device is applied all the electrodes have their magnetization aligned in the same direction.
By sweeping the magnetic field to positive values, the magnetization of the electrodes switches direction at their respective coercive fields, causing abrupt steps in the non-local resistance (red traces in Fig. \ref{fig:fig2}(a) and (b)).
Once the magnetization of all the electrodes again point in the same direction, the magnetic field is reversed and scanned from zero to negative values (blue traces in Fig. \ref{fig:fig2}(a) and (b)) showing a symmetric response in the non-local resistance.
Three states can be clearly identified: A, when the magnetization of all the electrodes are aligned; B, when one of the outer electrodes (1 or 6) switch its magnetization; and C, after the switch of one of the inner electrodes (2 or 4).
From the width of the contacts we can assume that the switching of the magnetization of the electrodes occurs at the order: 1, 6, 2 and 4.
Since we do not observe a switch due to one of the outer contacts, the configuration for each of the steps would be, in the order (1 2 4 6): A ($\uparrow \uparrow \uparrow \uparrow$), B ($\downarrow \uparrow \uparrow \downarrow$) and C ($\downarrow \downarrow \uparrow \downarrow$).
As illustrated in Fig. \ref{fig:fig2}(a) and (b), the magnitude and polarity of the switches in $R_{nl}$ are different for different values of $V_{bg}$.
In order to study how the size and sign of the switches in $R_{nl}$ change as a function of $V_{bg}$ we set the device at one of the magnetization configurations and record the non-local signal as a function of $V_{bg}$ at zero magnetic field.
The obtained $V_{bg}$ dependency of the $R_{nl}$ for each of the three states is shown in Fig. \ref{fig:fig2}(c).

On a close inspection of Fig. \ref{fig:fig2}(c), some peculiarities can be noted.
First, there is a clear structure that shows up for all three magnetization configurations.
This is specially evident around $V_{bg}\approx$0 V and can be attributed to a charge contribution to the non-local signal.
It can be also noted that $R_{nl}$ varies from negative to positive values in a wide range, from $\approx$ -300 to $\approx$ 100 $\Omega$.
Second, the values for $R_{nl}$ at different magnetization configuration do not keep a constant spacing between each other as a function of $V_{bg}$, but get modulated and even cross each other at a few points.
And finally, the values of $R_{nl}$ for the states A and B are centred around zero $\Omega$ and the ones corresponding to state C are centred aound $\approx$-50 $\Omega$.
This is what one would expect for the spin dependent signal considering the geometry of the device, since the device dimensions are smaller than $\lambda_{s}$ for both graphene regions in the injection and detection circuits \cite{PRB-Zaffalon}.
Taking as an example the injection circuit, when we drive a current through the electrodes in a parallel configuration, one of the electrodes will work as a spin up injector and the other as a spin up extractor.
For the case of contacts with equal polarization, the total spin accumulation created in the region is approximately zero.
On the other hand, when the electrodes are antiparallel, both will effectively inject spin up, creating then a finite spin accumulation \cite{PRB-Zaffalon}.
By reciprocity, a similar argument can be drawn for the detection circuit.
Finite element (see supporting information) of the device using conservative values for a contact polarization of P=10 \%, square resistance of the graphene $R_{sq}=1$ k$\Omega$ and $\lambda_{s}$= 1 $\mu$m show that a non-local spin valve signal in the order of $R_{nl}\approx$ 10 $\Omega$ is expected in the absence of quantum coherence effects.

Since our device is phase coherent, the non-local technique cannot fully exclude charge contributions.
Therefore, the detected $R_{nl}$ contains contributions from both the charge and the spin transmission through the constriction.
Indeed, as indicated by the arrows (and the oscillations between them) in Fig. \ref{fig:fig1}(b) and Fig. \ref{fig:fig2}(c) some similarities can be observed between the local and non-local signal.
To isolate the spin dependent part of the signal, we do the subtraction $\Delta R_{nl}^{B-A}=R_{nl}^{B}-R_{nl}^{A}$ and $\Delta R_{nl}^{A-C}=R_{nl}^{A}-R_{nl}^{C}$.
The values of $\Delta R_{nl}^{B-A}$ shows mainly the spin signal due to one the outer contacts, whereas $\Delta R_{nl}^{A-C}$ is mainly due to one of the inner contacts, which is the dominant contribution in the response of the spin-valve.
The back-gate dependences of $\Delta R_{nl}^{B-A}$ and $\Delta R_{nl}^{A-C}$ are shown in Fig. \ref{fig:fig2}(d).

When comparing Fig. \ref{fig:fig2}(c) and (d), $\Delta R_{nl}$ and $R_{nl}$ contain oscillations of similar magnitudes.
However, the spin polarization of the electrodes has still to be taken into account.
As discussed before, finite element simulations without the inclusion of quantum coherence effects and using typical values for our contacts\cite{Nature-Niko,NatPhys-Ivan,PRB-Mihai,PRB-Csaba,PRB-Thomas-FLG,PRB-Paul,NanoLett-Marcos} of $P=0.1$ give a spin valve signal in the same order as the average measured value.
Taking into account the efficiency of spin injection by the injector electrodes and the efficiency of spin detection by the detector electrodes, the non-local spin signal $R_{nl} \propto P^{2}$.
This implies that, in the case of $P\approx 1$, the spin dependent part on the non-local signal is about two orders of magnitude larger than the charge dependent part.

The larger spin contribution to the oscillations in the non-local signal can be explained by the fact that in order to non-locally observe the UCF we need to create a non-equilibrium situation (e.g. voltage bias).
In the case of the charge contribution to the fluctuations, we can expect that the non-equilibrium situation decays away from the current injection electrode on the scale of the phase coherence length.
Since the experiments are performed at a finite temperature, we also have to consider its limiting factor for the phase coherence.
The temperature can be taken into account by the thermal length $\lambda_{T}= \sqrt{\hbar D / k_{B} T} \approx $ 100 nm, where $\hbar$ is the reduced Planck's constant, $D$ the diffusion constant, $k_{B}$ the Boltzmann constant and $T$ the temperature.
This means that the non-equilibrium situation for the charge is maintained over only a few hundred nanometers.
However, for the spins a non-equilibrium situation is maintained by the spin accumulation which decays very slowly, in the order of $\lambda_{s}\approx$ 1 $\mu$m, allowing for the observation of coherent effects in the spin signal on longer length scales.

\begin{figure}[h]
    \centering
        \includegraphics[width=0.35\textwidth]{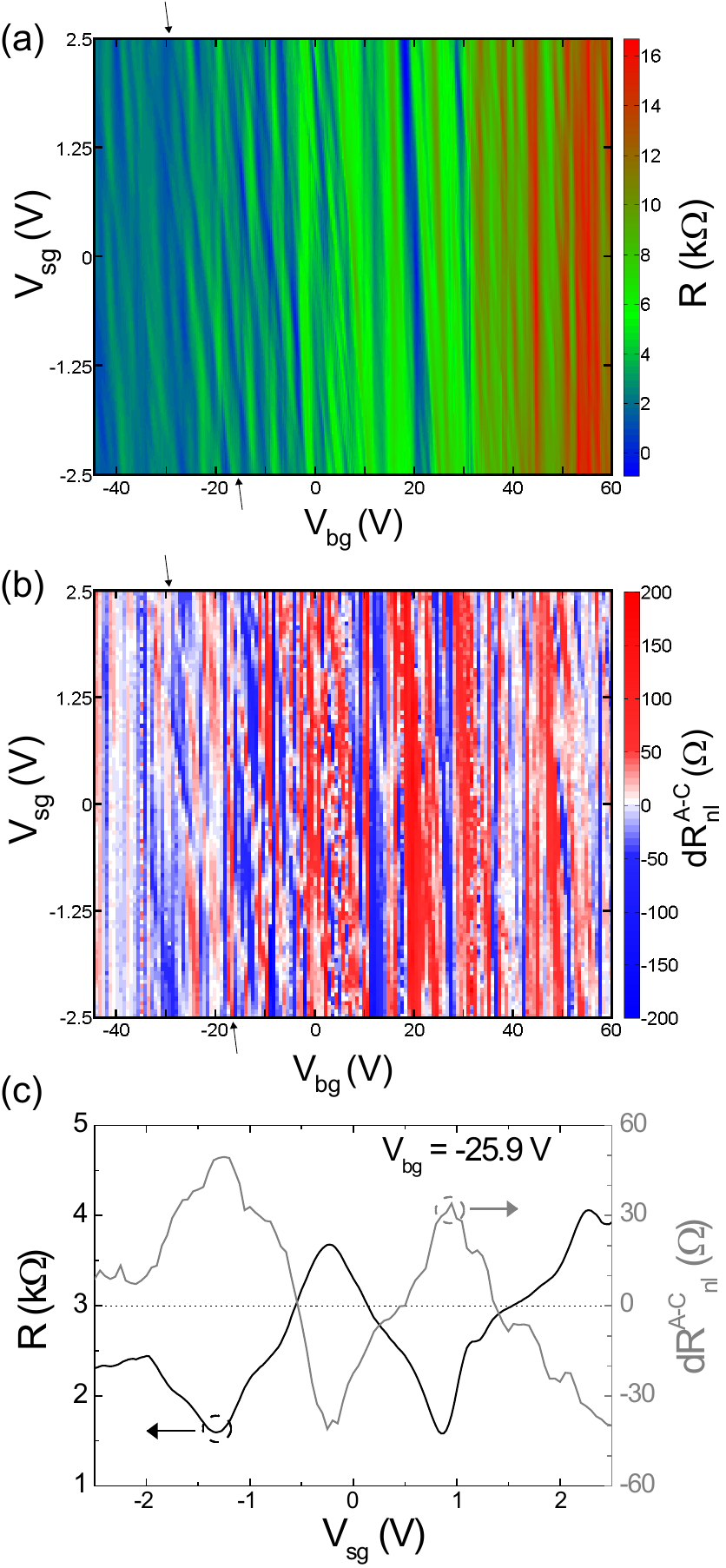}
    \caption{Side- and back-gate voltage dependence of the (a)local 4-terminal resistance and (b) spin-valve signal ($\Delta R^{A-C}_{nl}$). The arrows are guide to the eye to exemplify similar features in the local and non-local signal. (c) Local resistance (black) and non-local spin signal (grey) as a function of $V_{sg}$ for a fixed $V_{bg}=-25.9$ V. These traces are indicated in (a) and (b) by dashed lines.}
    \label{fig:fig3}
\end{figure}

In order to further investigate the modulation and sign reversal in the spin-valve signal we performed measurements on the local and non-local resistance as a function of both side- and back-gate voltages (Fig. \ref{fig:fig3}).
A careful look at Fig. \ref{fig:fig3}(a) reveals tilted line features showing that the fluctuations in resistance depends on both $V_{bg}$ and $V_{sg}$.
This is specially clear for $V_{bg}<0$ V.

Although the trends for the $R_{nl}$ are not as clear as those for the local signal, a few features below $V_{bg}=0$ V show a similar behaviour (see arrows on Fig. \ref{fig:fig3}(a) and (b) for comparison).
It can also be seen that by only changing the side-gate voltage the spin signal is not only strongly modulated, but also shows the sign reversal.
This is shown in \ref{fig:fig3}(b) by the colours blue (negative spin signal) and red (positive spin signal).
This effect is specially clear when we isolate a single trace in $V_{sg}$ from the scans of Fig.\ref{fig:fig3}(b) as shown in Fig. \ref{fig:fig3}(c) for a fixed $V_{bg}=-25.9$ V, indicated by a dashed line in Fig. \ref{fig:fig3}(a) and (b).
Given the local influence of the side-gate, this clearly indicates the major role of the constriction for the oscillations in the non-local spin signal.

The similarities between the local and non-local spin signal as a function of the side-gate (Fig. \ref{fig:fig3}(c)) strongly suggest that the inversion of the polarity of the spin signal is due to the UCF in the constriction.
This type of spin filtering effect was already demonstrated by Folk et al. in quantum dots\cite{Science-Folk}.
However, in order to prove that this effect can explain our results, a more detailed study on the dependence of the non-local signal on the spin accumulation is required.
The observation of such a strong modulation and sign reversal in the spin signal without the presence of magnetic fields demonstrates the potential of quantum coherent graphene structures for applications in future spintronic devices.

\section{Conclusions}
In conclusion we observed a strong oscillation in the non-local signal in a graphene nanostructure based spin-valve.
The oscillations in the non-local signal, attributed to UCF, showed to have spin and charge related contributions to the oscillations.
By changing the magnetization direction of our ferromagnetic electrodes we could separate the spin contribution to the signal and showed that, when the polarization of the electrodes is taken into account, the oscillations in the non-local signal due to spin is about two orders of magnitude higher than the charge.
We also showed that the non-local spin signal is not only strongly modulated but also reverses polarity.
Using a local gate to tune the carrier density in the constriction we demonstrated that the constriction is the main contributor to the oscillations in the non-local signal.
Our results indicate that the constriction can work as a spin filter device and show that graphene nanostructures have a great potential for future quantum spintronic applications.

\acknowledgement
This work was realized using NanoLab.NL (NanoNed) facilities and 
is part of the research program of the Foundation for Fundamental Research on Matter (FOM), 
which is part of the Netherlands Organization for Scientific Research (NWO).
We would like to acknowledge J. G. Holstein, H. M. de Roosz and B. Wolfs for technical support.

\newpage
\section{Supporting Information}

\subsection{Finite element calculations for the spin accumulation in the device}

\begin{figure}[h]
    \centering
        \includegraphics[width=0.65\textwidth]{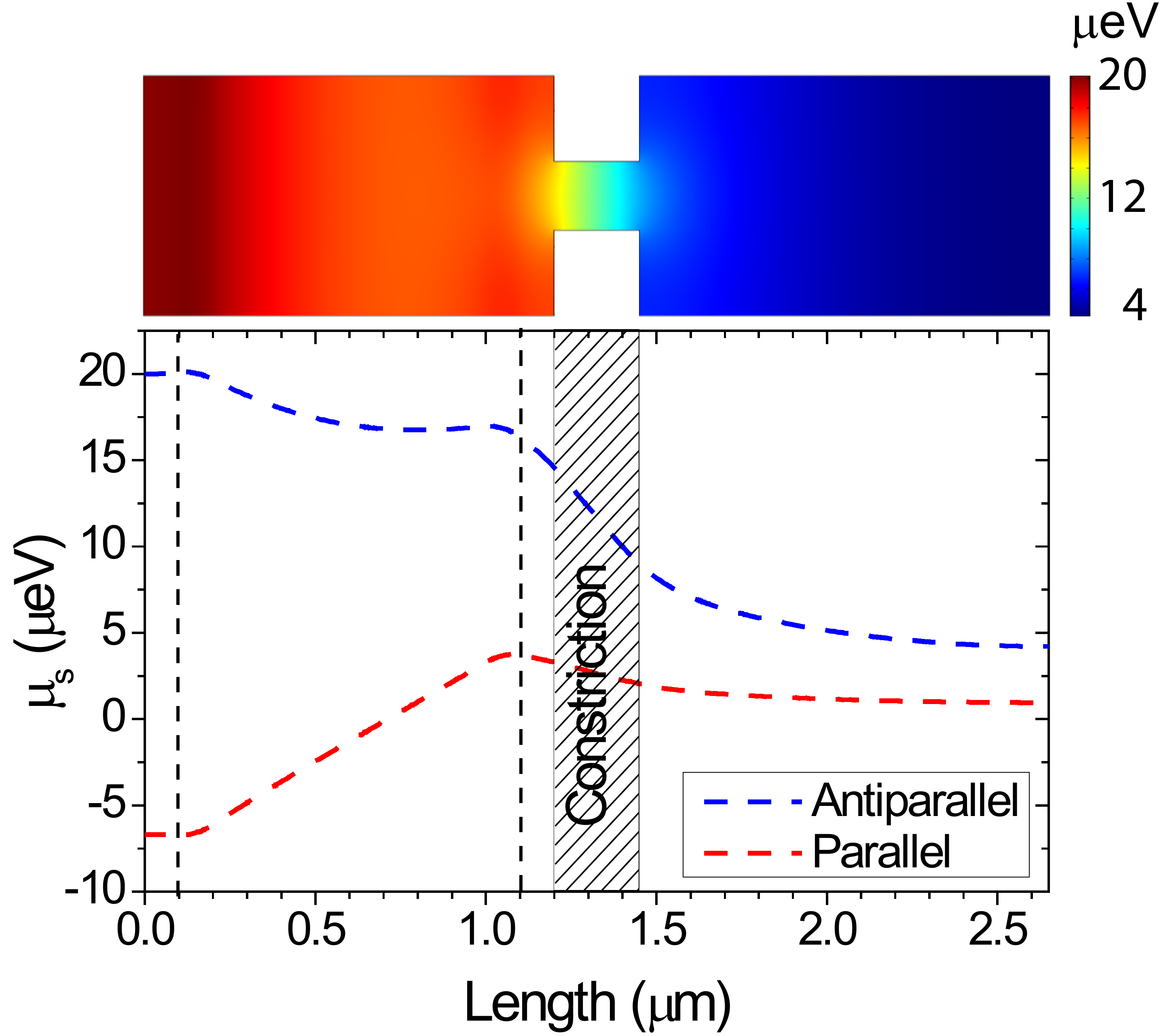}
    \caption{Finite element modelling of the classical spin accumulation in our device. On the top: device geometry with the location of the contacts shown by the dashed lines. The colour represents the spin accumulation according to the scale on the right for the case of anti-parallel alignment of the injection contacts. Bottom: Spin accumulation as a function of distance in the centre of the device for both parallel and anti-parallel configuration of the injection electrodes.}
    \label{fig:spinacc}
\end{figure}

In order to calculate the classical spin accumulation expected for the device we used finite element simulations as implemented in the software \textsc{COMSOL$^{\textregistered}$ MULTIPHYSICS}.
For this we solved the equations for spin diffusion: $\nabla^{2} \mu_{s} - \frac{\mu_{s}}{\lambda_{s}^{2}}=0$ for our confined device geometry with a current of 100 nA applied between two electrodes with spin polarization $P=10 \%$.
We assumed typical values for the square resistance of the graphene flake of $R_{sq}=1$ k$\Omega$ which is in the same order order of magnitude as in our experiment.
Furthermore, we assumed a conservative value for the spin relaxation length of $\lambda_{s}$ = 1 $\mu$m.
From the results of the simulation (shown in Fig. \ref{fig:spinacc}) we obtain a non-local spin valve signal of $\approx$ 15 $\Omega$ which is in the same order of magnitude of the average experimental value for the non-local spin signal as a function of back-gate voltage presented in the main text ($\Delta R_{nl}^{A-C} \approx $50 $\Omega$).
It is worth noting that this simulation uses a purely classical diffusion picture, which means that coherence effects are not included.
Although such a picture describes well the spin transport of previous experimental studied on non-local graphene spin-valves \cite{NatPhys-Ivan} and also the average observed non-local spin signal, it fails to explain the oscillations in the non-local signal as a function of gate voltage.

\subsection{Effect of the stray magnetic fields from the side-gate electrode}

\begin{figure}[h]
    \centering
        \includegraphics[width=0.65\textwidth]{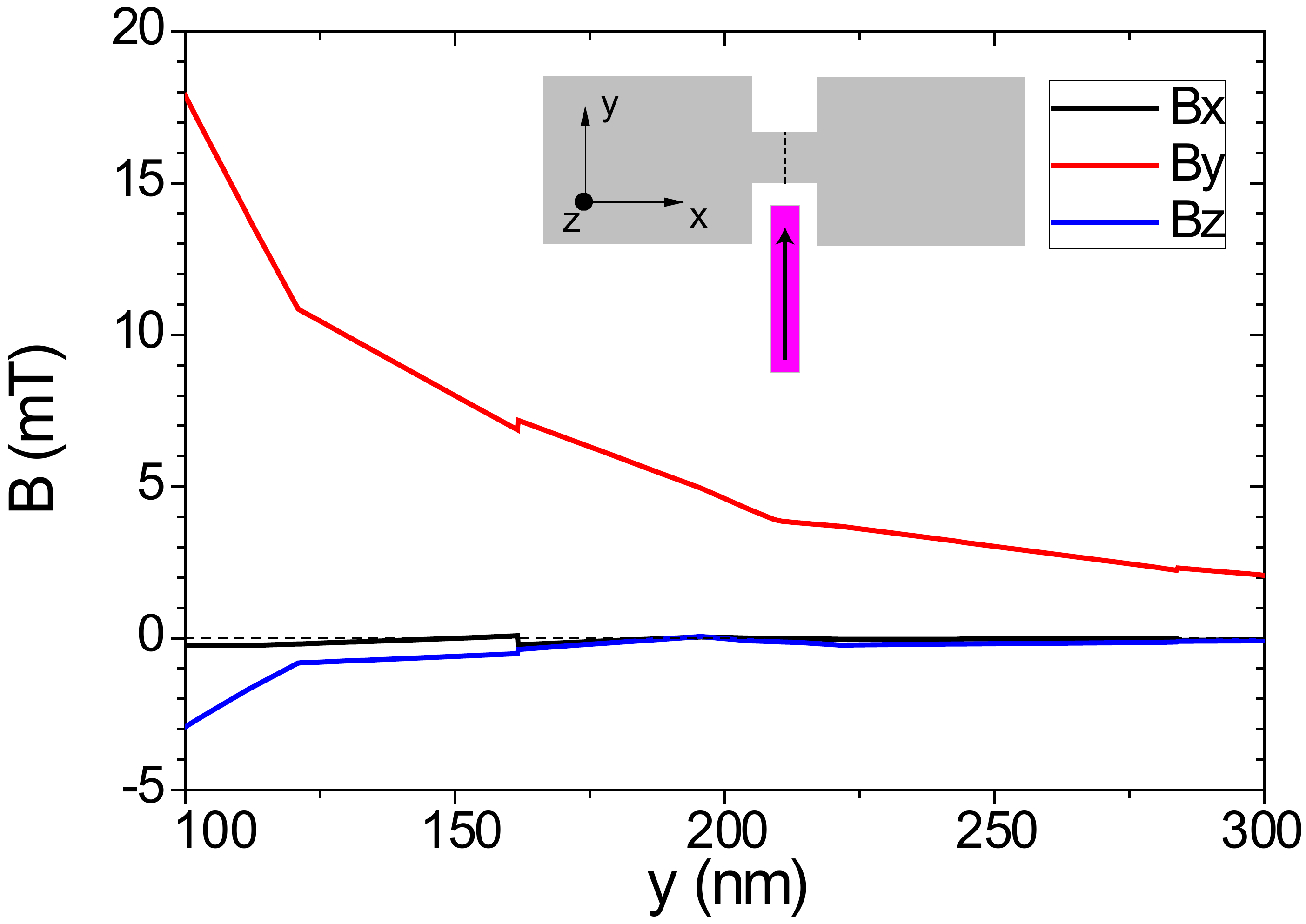}
    \caption{Components of the magnetic field in the x, y and z directions as a function of the distance for a cobalt bar of dimensions 0.1 $\times$ 10 $\times$ 0.035 $\mu$m$^{3}$. The results are shown for a line across the centre of the constriction as shown in the inset.}
    \label{fig:stray}
\end{figure}

In order to keep the contact interface as clean as possible we lowered the number of fabrication steps and fabricated the contacts and the side-gates at the same time.
Therefore, the side-gate electrodes are also magnetic.
To ensure that the stray magnetic fields arising from the side-gate electrode do not influence our results we performed finite element modelling of a cobalt bar of the same dimensions as the electrode in question.
The bar is assumed to be uniformly magnetized up to its end, which gives a maximum estimate of the stray magnetic fields.
In Fig. \ref{fig:stray} we show the results of the simulation in the region of the center of the constriction (100 nm $\leq$ y $\leq$ 300 nm).

As can be seen, the values for the out-of-plane component (z) is too small to create any orbital effect in the constriction.
For the in-plane component, the maximum field is about 20 mT.
This would create a Zeeman splitting of $E_{Z}=2.3$ $\mu$eV, which is much smaller than the thermal energy $E_{T}=361$ $\mu$eV.
Therefore we do not expect that the stray fields from the side-gate electrode would affect our findings.

To consider the effects of spin precession we have to take into account the out-of-plane component of the magnetic field $B_{z} \leq$ 3 mT which is too small to show any measurable effect in our measurements.

\bibliography{qis_bib}

\providecommand{\latin}[1]{#1}
\providecommand*\mcitethebibliography{\thebibliography}
\csname @ifundefined\endcsname{endmcitethebibliography}
  {\let\endmcitethebibliography\endthebibliography}{}
\begin{mcitethebibliography}{27}
\providecommand*\natexlab[1]{#1}
\providecommand*\mciteSetBstSublistMode[1]{}
\providecommand*\mciteSetBstMaxWidthForm[2]{}
\providecommand*\mciteBstWouldAddEndPuncttrue
  {\def\EndOfBibitem{\unskip.}}
\providecommand*\mciteBstWouldAddEndPunctfalse
  {\let\EndOfBibitem\relax}
\providecommand*\mciteSetBstMidEndSepPunct[3]{}
\providecommand*\mciteSetBstSublistLabelBeginEnd[3]{}
\providecommand*\EndOfBibitem{}
\mciteSetBstSublistMode{f}
\mciteSetBstMaxWidthForm{subitem}{(\alph{mcitesubitemcount})}
\mciteSetBstSublistLabelBeginEnd
  {\mcitemaxwidthsubitemform\space}
  {\relax}
  {\relax}

\bibitem[Huertas-Hernando \latin{et~al.}(2009)Huertas-Hernando, Guinea, and
  Brataas]{PRL-Brataas}
Huertas-Hernando,~D.; Guinea,~F.; Brataas,~A. \emph{Phys. Rev. Lett.}
  \textbf{2009}, \emph{103}, 146801\relax
\mciteBstWouldAddEndPuncttrue
\mciteSetBstMidEndSepPunct{\mcitedefaultmidpunct}
{\mcitedefaultendpunct}{\mcitedefaultseppunct}\relax
\EndOfBibitem
\bibitem[Zomer \latin{et~al.}(2012)Zomer, Guimar\~aes, Tombros, and van
  Wees]{PRB-Paul}
Zomer,~P.~J.; Guimar\~aes,~M. H.~D.; Tombros,~N.; van Wees,~B.~J. \emph{Phys.
  Rev. B} \textbf{2012}, \emph{86}, 161416\relax
\mciteBstWouldAddEndPuncttrue
\mciteSetBstMidEndSepPunct{\mcitedefaultmidpunct}
{\mcitedefaultendpunct}{\mcitedefaultseppunct}\relax
\EndOfBibitem
\bibitem[Vera-Marun \latin{et~al.}(2012)Vera-Marun, Ranjan, and van
  Wees]{NatPhys-Ivan}
Vera-Marun,~I.~J.; Ranjan,~V.; van Wees,~B.~J. \emph{Nat. Phys.} \textbf{2012},
  \emph{8}, 313\relax
\mciteBstWouldAddEndPuncttrue
\mciteSetBstMidEndSepPunct{\mcitedefaultmidpunct}
{\mcitedefaultendpunct}{\mcitedefaultseppunct}\relax
\EndOfBibitem
\bibitem[Guimarães \latin{et~al.}(2012)Guimarães, Veligura, Zomer, Maassen,
  Vera-Marun, Tombros, and van Wees]{NanoLett-Marcos}
Guimarães,~M. H.~D.; Veligura,~A.; Zomer,~P.~J.; Maassen,~T.;
  Vera-Marun,~I.~J.; Tombros,~N.; van Wees,~B.~J. \emph{Nano Letters}
  \textbf{2012}, \emph{12}, 3512--3517\relax
\mciteBstWouldAddEndPuncttrue
\mciteSetBstMidEndSepPunct{\mcitedefaultmidpunct}
{\mcitedefaultendpunct}{\mcitedefaultseppunct}\relax
\EndOfBibitem
\bibitem[J\'ozsa \latin{et~al.}(2009)J\'ozsa, Maassen, Popinciuc, Zomer,
  Veligura, Jonkman, and van Wees]{PRB-Csaba}
J\'ozsa,~C.; Maassen,~T.; Popinciuc,~M.; Zomer,~P.~J.; Veligura,~A.;
  Jonkman,~H.~T.; van Wees,~B.~J. \emph{Phys. Rev. B} \textbf{2009}, \emph{80},
  241403\relax
\mciteBstWouldAddEndPuncttrue
\mciteSetBstMidEndSepPunct{\mcitedefaultmidpunct}
{\mcitedefaultendpunct}{\mcitedefaultseppunct}\relax
\EndOfBibitem
\bibitem[Maassen \latin{et~al.}(2012)Maassen, Vera-Marun, Guimar\~aes, and van
  Wees]{PRB-Thomas-Contact}
Maassen,~T.; Vera-Marun,~I.~J.; Guimar\~aes,~M. H.~D.; van Wees,~B.~J.
  \emph{Phys. Rev. B} \textbf{2012}, \emph{86}, 235408\relax
\mciteBstWouldAddEndPuncttrue
\mciteSetBstMidEndSepPunct{\mcitedefaultmidpunct}
{\mcitedefaultendpunct}{\mcitedefaultseppunct}\relax
\EndOfBibitem
\bibitem[Maassen \latin{et~al.}(2011)Maassen, Dejene, Guimar\~aes, J\'ozsa, and
  van Wees]{PRB-Thomas-FLG}
Maassen,~T.; Dejene,~F.~K.; Guimar\~aes,~M. H.~D.; J\'ozsa,~C.; van Wees,~B.~J.
  \emph{Phys. Rev. B} \textbf{2011}, \emph{83}, 115410\relax
\mciteBstWouldAddEndPuncttrue
\mciteSetBstMidEndSepPunct{\mcitedefaultmidpunct}
{\mcitedefaultendpunct}{\mcitedefaultseppunct}\relax
\EndOfBibitem
\bibitem[Han and Kawakami(2011)Han, and Kawakami]{PRL-Kawakami}
Han,~W.; Kawakami,~R.~K. \emph{Phys. Rev. Lett.} \textbf{2011}, \emph{107},
  047207\relax
\mciteBstWouldAddEndPuncttrue
\mciteSetBstMidEndSepPunct{\mcitedefaultmidpunct}
{\mcitedefaultendpunct}{\mcitedefaultseppunct}\relax
\EndOfBibitem
\bibitem[Han \latin{et~al.}(2012)Han, McCreary, Pi, Wang, Li, Wen, Chen, and
  Kawakami]{JMMM-Kawakami}
Han,~W.; McCreary,~K.; Pi,~K.; Wang,~W.; Li,~Y.; Wen,~H.; Chen,~J.;
  Kawakami,~R. \emph{Journal of Magnetism and Magnetic Materials}
  \textbf{2012}, \emph{324}, 369 -- 381\relax
\mciteBstWouldAddEndPuncttrue
\mciteSetBstMidEndSepPunct{\mcitedefaultmidpunct}
{\mcitedefaultendpunct}{\mcitedefaultseppunct}\relax
\EndOfBibitem
\bibitem[Han \latin{et~al.}(2012)Han, Chen, Wang, McCreary, Wen, Swartz, Shi,
  and Kawakami]{NanoLett-Kawakami}
Han,~W.; Chen,~J.-R.; Wang,~D.; McCreary,~K.~M.; Wen,~H.; Swartz,~A.~G.;
  Shi,~J.; Kawakami,~R.~K. \emph{Nano Letters} \textbf{2012}, \emph{12},
  3443--3447\relax
\mciteBstWouldAddEndPuncttrue
\mciteSetBstMidEndSepPunct{\mcitedefaultmidpunct}
{\mcitedefaultendpunct}{\mcitedefaultseppunct}\relax
\EndOfBibitem
\bibitem[Swartz \latin{et~al.}(2013)Swartz, Chen, McCreary, Odenthal, Han, and
  Kawakami]{PRB-Kawakami}
Swartz,~A.~G.; Chen,~J.-R.; McCreary,~K.~M.; Odenthal,~P.~M.; Han,~W.;
  Kawakami,~R.~K. \emph{Phys. Rev. B} \textbf{2013}, \emph{87}, 075455\relax
\mciteBstWouldAddEndPuncttrue
\mciteSetBstMidEndSepPunct{\mcitedefaultmidpunct}
{\mcitedefaultendpunct}{\mcitedefaultseppunct}\relax
\EndOfBibitem
\bibitem[Kozikov \latin{et~al.}(2012)Kozikov, Horsell, McCann, and
  Fal'ko]{PRB-Falko}
Kozikov,~A.~A.; Horsell,~D.~W.; McCann,~E.; Fal'ko,~V.~I. \emph{Phys. Rev. B}
  \textbf{2012}, \emph{86}, 045436\relax
\mciteBstWouldAddEndPuncttrue
\mciteSetBstMidEndSepPunct{\mcitedefaultmidpunct}
{\mcitedefaultendpunct}{\mcitedefaultseppunct}\relax
\EndOfBibitem
\bibitem[Chen \latin{et~al.}(2010)Chen, Bae, Chialvo, Dirks, Bezryadin, and
  Mason]{JOP-Mason}
Chen,~Y.-F.; Bae,~M.-H.; Chialvo,~C.; Dirks,~T.; Bezryadin,~A.; Mason,~N.
  \emph{Journal of Physics: Condensed Matter} \textbf{2010}, \emph{22},
  205301\relax
\mciteBstWouldAddEndPuncttrue
\mciteSetBstMidEndSepPunct{\mcitedefaultmidpunct}
{\mcitedefaultendpunct}{\mcitedefaultseppunct}\relax
\EndOfBibitem
\bibitem[Lundeberg \latin{et~al.}(2013)Lundeberg, Yang, Renard, and
  Folk]{PRL-Folk-2013}
Lundeberg,~M.~B.; Yang,~R.; Renard,~J.; Folk,~J.~A. \emph{Phys. Rev. Lett.}
  \textbf{2013}, \emph{110}, 156601\relax
\mciteBstWouldAddEndPuncttrue
\mciteSetBstMidEndSepPunct{\mcitedefaultmidpunct}
{\mcitedefaultendpunct}{\mcitedefaultseppunct}\relax
\EndOfBibitem
\bibitem[Tombros \latin{et~al.}(2007)Tombros, Jozsa, Popinciuc, Jonkman, and
  van Wees]{Nature-Niko}
Tombros,~N.; Jozsa,~C.; Popinciuc,~M.; Jonkman,~H.~T.; van Wees,~B.~J.
  \emph{Nature} \textbf{2007}, \emph{448}, 571--574\relax
\mciteBstWouldAddEndPuncttrue
\mciteSetBstMidEndSepPunct{\mcitedefaultmidpunct}
{\mcitedefaultendpunct}{\mcitedefaultseppunct}\relax
\EndOfBibitem
\bibitem[Popinciuc \latin{et~al.}(2009)Popinciuc, J\'ozsa, Zomer, Tombros,
  Veligura, Jonkman, and van Wees]{PRB-Mihai}
Popinciuc,~M.; J\'ozsa,~C.; Zomer,~P.~J.; Tombros,~N.; Veligura,~A.;
  Jonkman,~H.~T.; van Wees,~B.~J. \emph{Phys. Rev. B} \textbf{2009}, \emph{80},
  214427\relax
\mciteBstWouldAddEndPuncttrue
\mciteSetBstMidEndSepPunct{\mcitedefaultmidpunct}
{\mcitedefaultendpunct}{\mcitedefaultseppunct}\relax
\EndOfBibitem
\bibitem[Folk \latin{et~al.}(2001)Folk, Patel, Birnbaum, Marcus, Duru\"oz, and
  Harris]{PRL-Folk}
Folk,~J.~A.; Patel,~S.~R.; Birnbaum,~K.~M.; Marcus,~C.~M.; Duru\"oz,~C.~I.;
  Harris,~J.~S. \emph{Phys. Rev. Lett.} \textbf{2001}, \emph{86},
  2102--2105\relax
\mciteBstWouldAddEndPuncttrue
\mciteSetBstMidEndSepPunct{\mcitedefaultmidpunct}
{\mcitedefaultendpunct}{\mcitedefaultseppunct}\relax
\EndOfBibitem
\bibitem[Folk \latin{et~al.}(2003)Folk, Potok, Marcus, and
  Umansky]{Science-Folk}
Folk,~J.~A.; Potok,~R.~M.; Marcus,~C.~M.; Umansky,~V. \emph{Science}
  \textbf{2003}, \emph{299}, 679--682\relax
\mciteBstWouldAddEndPuncttrue
\mciteSetBstMidEndSepPunct{\mcitedefaultmidpunct}
{\mcitedefaultendpunct}{\mcitedefaultseppunct}\relax
\EndOfBibitem
\bibitem[Hauptmann \latin{et~al.}(2008)Hauptmann, Paaske, and
  Lindelof]{NatPhys-Lindelof}
Hauptmann,~J.~R.; Paaske,~J.; Lindelof,~P.~E. \emph{Nat. Phys.} \textbf{2008},
  \emph{4}, 373--376\relax
\mciteBstWouldAddEndPuncttrue
\mciteSetBstMidEndSepPunct{\mcitedefaultmidpunct}
{\mcitedefaultendpunct}{\mcitedefaultseppunct}\relax
\EndOfBibitem
\bibitem[Sahoo \latin{et~al.}(2005)Sahoo, Kontos, Furer, Hoffmann, Gräber,
  Cottet, and Schönenberger]{NatPhys-Schonenberger}
Sahoo,~S.; Kontos,~T.; Furer,~J.; Hoffmann,~C.; Gräber,~M.; Cottet,~A.;
  Schönenberger,~C. \emph{Nat. Phys.} \textbf{2005}, \emph{1}, 99--102\relax
\mciteBstWouldAddEndPuncttrue
\mciteSetBstMidEndSepPunct{\mcitedefaultmidpunct}
{\mcitedefaultendpunct}{\mcitedefaultseppunct}\relax
\EndOfBibitem
\bibitem[Lundeberg and Folk(2009)Lundeberg, and Folk]{NatPhys-Folk}
Lundeberg,~M.~B.; Folk,~J.~A. \emph{Nat. Phys.} \textbf{2009}, \emph{5},
  894--897\relax
\mciteBstWouldAddEndPuncttrue
\mciteSetBstMidEndSepPunct{\mcitedefaultmidpunct}
{\mcitedefaultendpunct}{\mcitedefaultseppunct}\relax
\EndOfBibitem
\bibitem[Man \latin{et~al.}(2006)Man, Wever, and Morpurgo]{PRB-Mopurgo}
Man,~H.~T.; Wever,~I. J.~W.; Morpurgo,~A.~F. \emph{Phys. Rev. B} \textbf{2006},
  \emph{73}, 241401\relax
\mciteBstWouldAddEndPuncttrue
\mciteSetBstMidEndSepPunct{\mcitedefaultmidpunct}
{\mcitedefaultendpunct}{\mcitedefaultseppunct}\relax
\EndOfBibitem
\bibitem[Gunnarsson \latin{et~al.}(2008)Gunnarsson, Trbovic, and
  Sch\"onenberger]{PRB-Schonenberger}
Gunnarsson,~G.; Trbovic,~J.; Sch\"onenberger,~C. \emph{Phys. Rev. B}
  \textbf{2008}, \emph{77}, 201405\relax
\mciteBstWouldAddEndPuncttrue
\mciteSetBstMidEndSepPunct{\mcitedefaultmidpunct}
{\mcitedefaultendpunct}{\mcitedefaultseppunct}\relax
\EndOfBibitem
\bibitem[Beenakker and van Houten(1991)Beenakker, and van Houten]{SSP-Beenaker}
Beenakker,~C.; van Houten,~H. In \emph{Semiconductor Heterostructures and
  Nanostructures}; Ehrenreich,~H., Turnbull,~D., Eds.; Solid State Physics;
  Academic Press, 1991; Vol.~44; pp 1 -- 228\relax
\mciteBstWouldAddEndPuncttrue
\mciteSetBstMidEndSepPunct{\mcitedefaultmidpunct}
{\mcitedefaultendpunct}{\mcitedefaultseppunct}\relax
\EndOfBibitem
\bibitem[Cho \latin{et~al.}(2007)Cho, Chen, and Fuhrer]{APL-Fuhrer}
Cho,~S.; Chen,~Y.-F.; Fuhrer,~M.~S. \emph{Applied Physics Letters}
  \textbf{2007}, \emph{91}, 123105\relax
\mciteBstWouldAddEndPuncttrue
\mciteSetBstMidEndSepPunct{\mcitedefaultmidpunct}
{\mcitedefaultendpunct}{\mcitedefaultseppunct}\relax
\EndOfBibitem
\bibitem[Zaffalon and van Wees(2005)Zaffalon, and van Wees]{PRB-Zaffalon}
Zaffalon,~M.; van Wees,~B.~J. \emph{Phys. Rev. B} \textbf{2005}, \emph{71},
  125401\relax
\mciteBstWouldAddEndPuncttrue
\mciteSetBstMidEndSepPunct{\mcitedefaultmidpunct}
{\mcitedefaultendpunct}{\mcitedefaultseppunct}\relax
\EndOfBibitem
\end{mcitethebibliography}

\end{document}